\documentclass[aps,prl,reprint,floatfix,amsmath,amssymb]{revtex4-1}

\usepackage{epsfig}
\usepackage{graphicx}
\usepackage{dcolumn}
\usepackage{bm}
\usepackage[colorlinks=true,dvipdfm]{hyperref}
\usepackage{amssymb}

\begin{document}
\title{Universal scaling in first-order phase transitions mixed with nucleation and growth}

\author{Fan Zhong}
\affiliation{State Key Laboratory of Optoelectronic Materials and
Technologies and School of Physics, Sun Yat-sen
University, Guangzhou 510275, People's Republic of China}

\date{\today}

\begin{abstract}
Matter exhibits phases and their transitions. These transitions are classified as first-order phase transitions (FOPTs) and continuous ones. While the latter has a well-established theory of the renormalization group, the former is only qualitatively accounted for by classical theories of nucleation, since their predictions often disagree with experiments by orders of magnitude. A theory to integrate FOPTs into the framework of the renormalization-group theory has been proposed but seems to contradict with extant wisdom and lacks numerical evidence. Here we show that universal hysteresis scaling as predicted by the renormalization-group theory emerges unambiguously when the theory is combined intimately with the theory of nucleation and growth in the FOPTs of the paradigmatic two-dimensional Ising model driven by a linearly varying externally applied field below its critical point. This not only provides a new method to rectify the nucleation theories, but also unifies the theories for both classes of transitions and FOPTs can be studied using universality and scaling similar to their continuous counterpart.
\end{abstract}

\maketitle

Matter as a many-body system exists in various phases and/or their coexistence and its diversity comes from phase changes. It thus exhibits just phases and their transitions. These transitions are classified as first-order phase transitions (FOPTs) and continuous ones~\cite{Fisher67}, the latter including second and higher orders. Whereas the phases can be studied by a well-established framework and the continuous phase transitions have a well-established theory of the renormalization group (RG) that has predicted precise results in good agreement with experiments~\cite{Barmatz}, the FOPTs gain a different status in statistical physics.

FOPTs proceed through either nucleation and growth or spinodal decomposition~\cite{Gunton83,Bray,Binder2}. Although classical theories of nucleation~\cite{Becker,Becker1,Becker2,Zeldovich,books,books1,books2,Oxtoby92,Oxtoby921,Oxtoby922,Oxtoby923} and growth~\cite{Avrami,Avrami1,Avrami2} correctly account for the qualitative features of a transition, an agreement in the nucleation rate of even several orders of magnitude between theoretical predictions and experimental and numerical results is regarded as a feat~\cite{Oxtoby92,Oxtoby921,Oxtoby922,Oxtoby923,Filion,Filion1,Filion2}. A lot of improvements have thus been proposed and tested in the two-dimensional (2D) Ising model whose exact solution is available. In a multidroplet regime~\cite{Rikvold94} in which many droplets nucleate and grow, by combining with Avrami's growth law~\cite{Avrami,Avrami1,Avrami2}, a field-theoretically-corrected nucleation theory~\cite{Langer67,Gunther,Harris84,Prestipino} was shown to produce quite well---with only one adjustable parameter---the results of hysteresis loop areas obtained from Monte Carlo simulations at temperatures below the critical temperature $T_c$ even in the case of a sinusoidally varying applied external field~\cite{Sides98,Sides99,Ramos,Zhongc}.

However, it is well-known that classical nucleation theories are not applicable in spinodal decompositions in which the critical droplet for nucleation is of the size of the lattice constant and thus no nucleation is needed~\cite{Gunton83}. In contrary to the mean-field case, for systems with short-range interactions, although sharply defined spinodals that divide the two regimes of the apparently different dynamic mechanisms do not exist~\cite{Gunton83,Bray,Binder2}, one can nevertheless assume existence of fluctuation shifted underlying spinodals called ``instability'' points. Expanding around them below $T_c$ of a usual $\phi^4$ theory for critical phenomena then results in a $\phi^3$ theory for the FOPT due to the lack of the up--down symmetry in the expansion~\cite{Zhongl05,zhong16}. An RG theory for the FOPT can then be set up in parallel to that for the critical phenomena, giving rise to universality and dynamic scaling characterized by analogous ``instability'' exponents. The primary qualitative difference is that the nontrivial fixed points of such a theory are imaginary in values and are thus usually considered to be unphysical, though the instability exponents are real. Yet, counter-intuitively, imaginariness is physical in order for the $\phi^{3}$ theory to be mathematically convergent, since the system becomes unstable at the instability points upon renormalization and analytical continuation is necessary~\cite{Zhonge12}. Moreover, the degrees of freedom that need finite free energy costs for nucleation are coarse-grained away with the costs, indicating irrelevancy of nucleation to the scaling~\cite{Zhonge12}. Although no clear evidence of an overall power-law relationship was found for the magnetic hysteresis in a sinusoidally oscillating field in two dimensions~\cite{Thomas,Sides98,Sides99} in contrast to previous work~\cite{Char,Liang16}, recently, with properly logarithmic corrections, a dynamic scaling near a temperature other than the equilibrium transition point was again found numerically for the cooling FOPTs in the 2D Potts model~\cite{Pelissetto16}. However, no theoretical explanation was offered to the scaling~\cite{Liang16}.

Here, we propose an idea that the instability point is reached when the time scale of the nucleation and growth matches that of the driving arising from the linearly varying field $H$ with a rate $R$. Integrating the theory of nucleation and growth with the $\phi^3$ RG theory of scaling for FOPTs, we are then able to construct a finite-time scaling form for the magnetization. It is found to describe remarkably well the numerical simulations of the 2D Ising model with universal instability exponents and scaling functions for two simulated temperatures below $T_c$ after allowing for a single additional universal logarithmic factor. Because the scaling form contains all essential elements of nucleation and growth including the Boltzmann factor that is the origin of the large discrepancy between nucleation theories and experiments, the scaling provides a method to rectify it. More importantly, our results offer unambiguous evidence for the $\phi^3$ theory and thus one can study the universality and scaling of FOPTs similar to their continuous counterpart.

Crucial in our analysis is the theory of finite-time scaling~\cite{Gong,Gong1,Huang,Feng}, whose essence is a constant finite time scale $t_R=\zeta_RR^{-z/r}$ arising from the linear driving, where $z$ and $r$ are dynamic instability exponents and $\zeta_R$ denotes the proportional coefficient independent on $R$. This single externally imposed time scale enables one to probe effectively a process in which a system takes a long time to equilibrate, as is the present case of nucleation and growth. This is because the system can then readily follow the short time instead of the long equilibration one. As a consequence, the system is controlled by the driving and exhibits finite-time scaling, similar to its spatial counterpart, finite-size scaling, in which a system has a smaller size than its correlation length. Moreover, even if crossover occurs when the equilibration time becomes shorter than, finite-time scaling can still well describe the situation~\cite{Huang,Feng}. By contrast, a sinusoidal driving has two controlling parameters, the amplitude and frequency, and thus complicates the process~\cite{Feng,Zhonge,Zhonge1}.

Let's start with the nucleation and growth of up spins in a sea of fluctuating down spins in the Ising model at a temperature $T$ below $T_c$. Upon applying a constant up field $H$ to the system, the magnetization $M$ at time $t$ is given by
\begin{equation}
M(H,T,t)=M_{\rm eq}(T)-2 M_{\rm eq}(T)\exp\left[-(t/t_0)^{d+1}\right],\label{mt}
\end{equation}
in the multidroplet regime~\cite{Rikvold94}, with a nucleation and growth time scale $t_0=\zeta_0H^{-(K+d)/(d+1)}\exp\{\Xi/[(d+1)H^{d-1}]\}$, where $M_{\rm eq}$ stands for the equilibrium spontaneous magnetization, $\zeta_0(T)=[\Omega_dv^dB/(d+1)]^{-1/(d+1)}$ is a temperature-dependent constant, $\Omega_d(T)$ is a shape factor, $B(T)$ is an adjustable parameter, $v(T)$ is the interface velocity of a growing droplet for a unit applied field in the Lifshitz-Allen-Cahn approximation~\cite{Lifshitz,Lifshitz1,Gunton83}, $K=3$ for the 2D kinetic Ising model, and $\Xi=\Omega_d\sigma_0^2/(2M_{\rm eq}k_{\rm B}T)$ with $k_{\rm B}$ being the Boltzmann constant and $\sigma_0$ the surface tension along a primitive lattice vector~\cite{Gunther,Rikvold94,Harris84}.

The central idea is that scaling emerges around the field $H_s$ that satisfies $t_0=t_R$. It divides regimes in which either nucleation and growth or the intrinsic fluctuations governed by the $\phi^3$ fixed point is dominant and thus is identified with the instability point of the theory, which was originally suggested to separate nucleation and growth from spinodal decomposition. This condition results in
\begin{equation}
H_s^{d-1}(-\ln R+\kappa \ln H_s+b)=r\Xi/[(d+1)z],\label{hs}
\end{equation}
with $\kappa\equiv r(K+d)/[z(d+1)]$ and $b\equiv r\ln(\zeta_R/\zeta_0)/z$, which is proportional to the ratio of the coefficients of the two time scales. For the field driven case, $r=z+\beta\delta/\nu$ with $\beta$, $\delta$, and $\nu$ being the instability exponents for the magnetization, the magnetic field, and the correlation length, respectively~\cite{Zhongl05,zhong16}. The corresponding $M_s$ is given by the magnetization at $H_s$ obtained from Eq.~(\ref{mt}) with $t$ replaced by $H/R$, i.e.,
\begin{equation}
M_s=M_{\rm eq}-2 M_{\rm eq}\exp\left[-(\zeta_0R)^{-(d+1)}H_s^{K+2d+1}e^{-\Xi/H_s^{d-1}}\right].\label{ms}
\end{equation}
Why this is called $M_s$ will become clear shortly.

Several remarks are in order. First, the instability points so obtained depend on the rate $R$. This is reasonable as they rely on the probing scales as previous studies have shown~\cite{Binder78,Kawasaki,Kaski}. Only in the case in which the first two terms on the left hand side of Eq.~(\ref{hs}) can be neglected can one arrive at a constant $H_s$. Second, as $R\rightarrow0$, $H_s\rightarrow0$ and $M_s\rightarrow M_{\rm eq}$, viz., the equilibrium transition point and magnetization, rather than the mean-field spinodal since the range of interactions is short. This is again reasonable in view of the new physical meaning of the instability point; because, at the equilibrium transition point, only nucleation and growth is possible though the transition may take a time longer than the age of the universe. Note that $M_s\rightarrow M_{\rm eq}$ instead of the initial state $-M_{\rm eq}$ as $R\rightarrow0$ since nucleation and growth have been considered. Third, if the last two terms on the left hand side of Eq.~(\ref{hs}) are neglected for sufficiently low $R$, one reaches~\cite{Thomas} $H_s\sim (-\ln R)^{-1/(d-1)}$, which vanishes only for so extremely low $R$ that is not feasible numerically or experimentally~\cite{Sides99,Zhongc}. As a result, $H_s$ is always finite practically~\cite{Zhonge}. Fourth, the recently found logarithmic time factor~\cite{Pelissetto16} should be an approximated form of~(\ref{hs}) as the scaling found there is peculiar in that the normalized energy is not rescaled both in curve crossing and in scaling collapses different from usual scaling in critical phenomena~\cite{cumulant}. So should those found numerically in Ref.~\cite{Zhongc}.

In the $\phi^3$ theory, near the instability point, scaling exists similar to critical phenomena. Under a linearly varying external field, the finite-time scaling form is~\cite{Zhongl05,zhong16},
\begin{equation}
M(H,T,R)-M_s=R^{\beta/r\nu}f\left((H-H_s)R^{-\beta\delta/r\nu}\right),\label{ftsu}
\end{equation}
where $f$ is a universal scaling function. A salient feature is the finite $H_s$ and $M_s$. Now, because the instability points are determined by Eqs.~(\ref{hs}) and (\ref{ms}), it is then natural to postulate that the scaling form changes to
\begin{equation}
Y(H,T,R)= R^{\beta/r\nu}f\left(XR^{-\beta\delta/r\nu}(-\ln R)^{-3/2}\right),\label{ftsf}
\end{equation}
with $X(H,T,R)\equiv H^{d-1}(-\ln R+\kappa \ln H+b)-r\Xi/[(d+1)z]$ and $Y(H,T,R)\equiv M(H,T,R)-M_{\rm eq}+2 M_{\rm eq}\exp\{-\zeta_0^{-(d+1)}H^{K+2d+1}R^{-(d+1)}e^{-\Xi/H^{d-1}}\}$. In~(\ref{ftsf}), we have included a logarithmic factor with an exponent $-3/2$. It may stem from either the $\phi^3$ theory or the neglected higher order terms in the nucleation rate~\cite{Gunther}. At present, we have no definite theory to explain it. However, we find that this single factor is sufficient for good scaling collapses for both temperatures we simulate and is thus universal for at least the model in two dimensions too.

From the scaling form~(\ref{ftsf}), at $X=0$, one recovers naturally $H_s$ that obeys Eq.~(\ref{hs}). Yet, the magnetization now satisfies
\begin{equation}
M(H_s,T,R)=M_s(T,R)+ R^{\beta/r\nu}f(0)\label{ftsm}
\end{equation}
similar to the one obtained from~(\ref{ftsu}) though $M_s$ defined in Eq.~(\ref{ms}) is rate dependent. However, at $M=M_s$ or $Y=0$, one can only find $X|_{Y=0}=aR^{\beta\delta/r\nu}(-\ln R)^{3/2}$ for $f(a)=0$, different from the usual form $H|_{M=M_s}=H_s+aR^{\beta\delta/r\nu}$ obtained from Eq.~(\ref{ftsu}).

We employ the 2D Ising model to verify the scaling form~(\ref{ftsf}). It contains only two unknown parameters, $\zeta_0$ and $\zeta_R$ or $b$. Although $v$ and $B$ that define $\zeta_0$ were estimated in the model at $T=0.8T_c$~\cite{Sides99}, $B$ was found by adjusting it to match the data without considering scaling. We therefore regard $\zeta_0$ as an adjustable parameter. The other parameters, $\Xi$ and $K$, which crucially affect the nucleation and growth, are known for the same model. $M_{\rm eq}$, along with $\Xi$, $\Omega_d$, and $\sigma_0$, is even exactly known. For the universal exponents, $\delta$ is exactly known to be six~\cite{Cardy85}. This gives $\nu=-5/2$ using the exact result $\beta=1$ for the $\phi^3$ theory~\cite{zhong16}. The former two differ their three loop results by five percents or so. In contrast, $z$ is only estimated to two loop orders~\cite{zhong16}. Yet, it determines almost everything as seen from Eqs.~(\ref{hs}) and (\ref{ftsf}). So, we have to adjust it to find the best results.

The procedure is as follows. Given a $z$, we guess a value of $b$ and solve out $H_s$ from Eq.~(\ref{hs}), find the corresponding $M(H_s)$ for a series of $R$ from the simulated magnetization curves, and then fit them according to Eqs.~(\ref{ftsm}) and (\ref{ms}). The correct $b$ must lead to the right $\beta/r\nu$ for the given $z$ or to $d + 1$ if the power of $R$ in the exponent instead of $\beta/r\nu$ is regarded as a parameter to be fitted. The fitted out $\zeta_0$ can then be plugged in Eq.~(\ref{ftsf}) to verify the results. Remarkably, the two time scale coefficients can even be found.

\begin{figure*}
\centerline{\includegraphics[width=\linewidth]{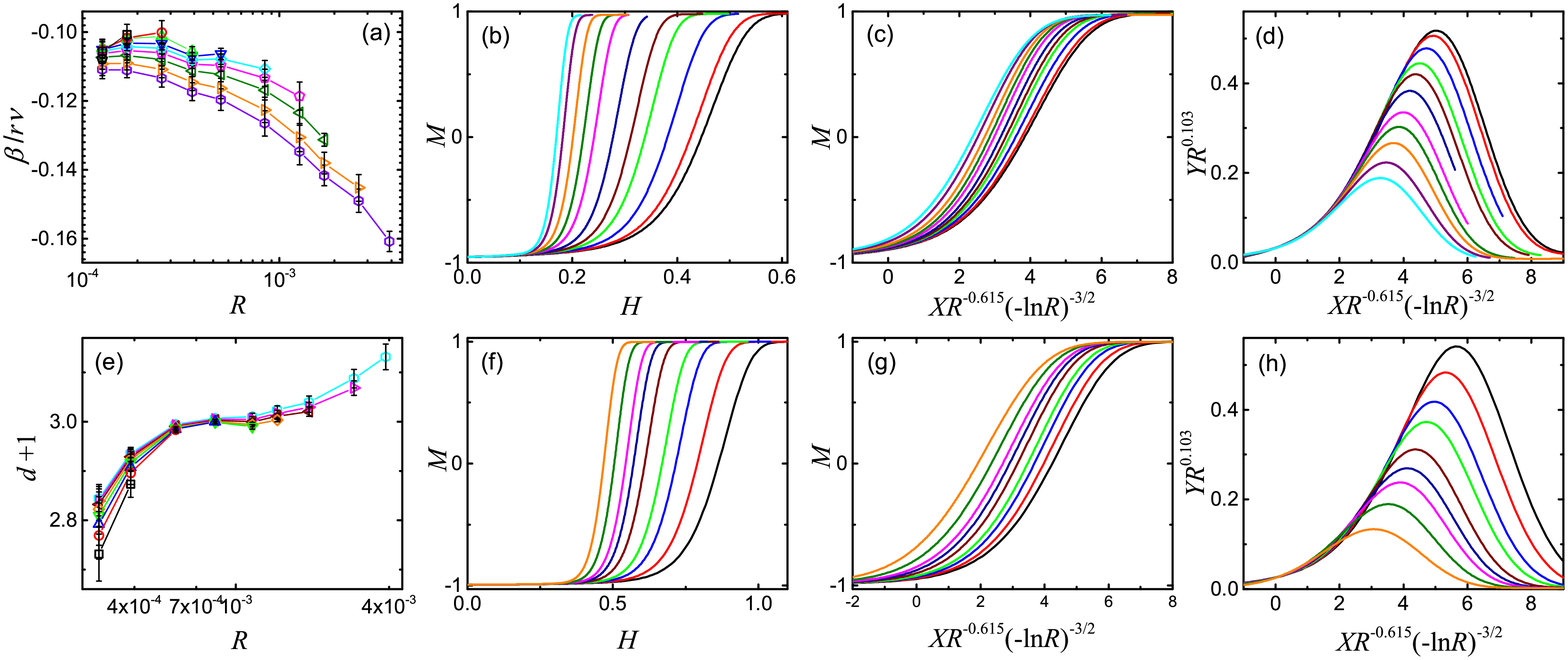}}
\caption{\label{rgt}(Color online) Fitting to the theory. (a) to (d) are results for $T=0.8T_c$ and (e) to (h) for  $T=1/0.735\approx0.6T_c$. (a) and (e) are $\beta/r\nu$ and $d +1$, respectively, fitted out from Eqs.~(\ref{ftsm}) and (\ref{ms}). For each curve, starting from the rightmost data point that represents the fit to the $R$ it stands and five others which are larger than it, each connected successive point denotes the fit of its $R$ and all the foregoing larger ones. The rates that give rise to approximately the correct exponents $-0.103$ and $3$ are thus chosen to be the five data from the sixth curve counting from the rightmost end point plus the five larger rates, adding up to ten rates ranging from about $0.00361$ to $0.000 126$ for $T = 0.8T_c$, and the three data from the fourth curve from the right plus the five larger rates, totally eight rates from $0.008 30$ to $0.000 830$ for $T \approx 0.6T_c$. The magnetization curves $M$ versus $H$ for these chosen rates are shown in (b) and (f). We have also shown one more large rate at $0.004 21$ (the fifth rate from the right in (b)) and one more small rate at $0.000 581$ (the seven rate from the right in (f)) for the two temperatures. As $R$ decreases, the curves shift to the left. (d) and (h) display the rescaled curves of all curves in (b) and (f), respectively. (c) and (g) depict the curves with only $H$ being rescaled in order to see to what extend of the original curves the collapsed parts correspond. Identical colors represent curves of identical rates for the same temperature only. $b=2.82$ and $0.77$ for $T = 0.8T_c$ and $T \approx 0.6T_c$, respectively. The exact values of $M_{\rm eq}$ and $\Xi$ are $0.954 411$ and $0.506 192$ for $T = 0.8T_c$, and $0.992 879$ and $2.202 925$ for $T \approx 0.6T_c$, respectively, while the fitted $\zeta_0$ is $36.3$ and $59.9$ for the two temperatures. Lines connecting symbols in (a) and (e) are only a guide to the eye.}
\end{figure*}

Figure~\ref{rgt} shows the results for $z = 1.5$. We use $\beta/r\nu$ and $d + 1$ to demonstrate the results of the fits to Eqs.~(\ref{ftsm}) and (\ref{ms}) for two different temperatures. The other ones that are not displayed show similar behaviour. One sees from \ref{rgt}(a) and \ref{rgt}(e) that, as the data of large rates $R$ are omitted in the fits, the exponents approach $-0.103$ and $3$ correctly. In \ref{rgt}(e), including of smaller rates again drives the exponent away from $3$, to which we shall come back later on. Using the fitted results of $\zeta_0$ and the ranges of rates that produce the correct exponents, we rescale the magnetization curves shown in \ref{rgt}(b) and \ref{rgt}(f) according to the scaling form~(\ref{ftsf}) and plot the results in \ref{rgt}(d) and \ref{rgt}(h). The peaks of the rescaled curves stem from the competition between $M$ and the part of nucleation and growth in $Y$ and lie in the late stages of the transition as seen in \ref{rgt}(c) and \ref{rgt}(g). One sees that the rescaled curves collapse onto each other almost perfectly even relatively far away from the instability point at $X = 0$ where the $\phi^3$ theory is developed~\cite{Zhongl05,zhong16} and even for rates beyond. This strongly validates the scaling form.

Similar scaling collapses appear for $z$ bigger than $1.5$ and even up to $2.5$ plus. We choose $1.5$ because the scaling functions for the two temperatures are nearly parallel. This can be seen in Figs.~\ref{rgt}(d) and \ref{rgt}(h), where identical scales are employed. The two rescaled curves only displace with each other by less than $0.01$ in $f(0)$.  This slight difference may result from the neglected higher order terms in the nucleation rate~\cite{Gunther}, which may also be a source of the extra logarithmic factor as mentioned. We note, however, that without the logarithmic factor, the rescaled curves already cross at $X = 0$ and $Y(H_s,T,R)R^{-\beta/r\nu}=f(0)$ almost perfectly in agreement with Eq.~(\ref{ftsm})) as demonstrated in Figs.~\ref{rgt}(a) and \ref{rgt}(e). This indicates that at least the predominant contribution of the nucleation and growth has been taken into account in the present theory.

\begin{figure}[b]
\centerline{\includegraphics[width=\linewidth]{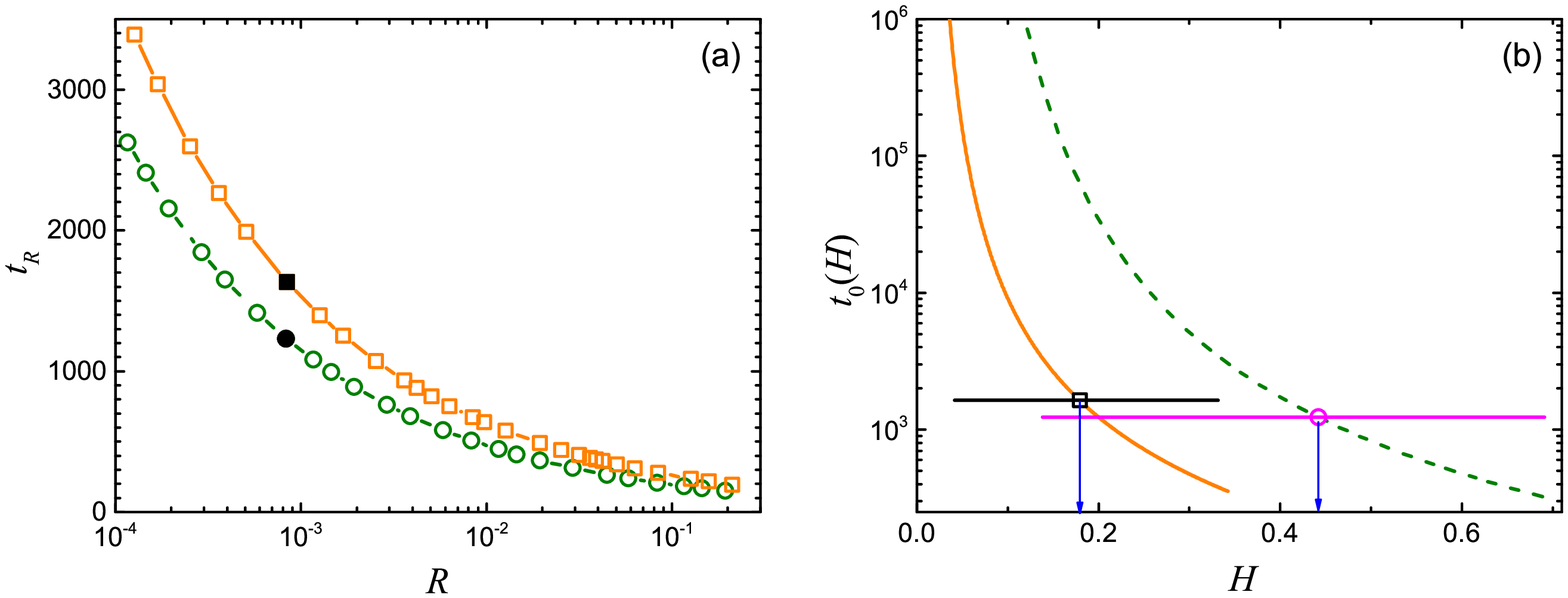}}
\caption{\label{trt0}(Color online) (a) The driving time scale $t_R$ versus $R$ and (b) the nucleation and growth time scale $t_0$ versus the field $H$ for the two nearly identical rates (filled black symbols) in (a) at $T = 0.8T_c$ (squares/orange) and $T \approx 0.6T_c$ (circles/olive). The two horizontal lines are $t_0(H_s) = t_R$, corresponding to the two filled symbols in (a). The two vertical arrows separate the intrinsic fluctuation regime on the left from the nucleation and growth regime on the right. Lines connecting symbols are only a guide to the eye.}
\end{figure}
Figure~\ref{trt0} displays different time scales. From its definition, $t_R$ (or $t_0$) increases as $R$ (or $H$) decreases and diverges as $R$ (or $H$) vanishes. However, $t_0$ climbs up exponentially fast than $t_R$ as can be seen in the figure. This implies that there exists always a finite-time regime in which the driving dominates the dynamics no matter how small the driving rate is. From the dependences of $t_R$ and $t_0$ on $R$ and $H$, it is evident that $H_s$ increases with $R$. As a consequence, large rates drive the transition to take place at large fields as exemplified in Figs.~\ref{rgt}(b) and (f). In addition, because the free-energy cost for nucleation increases significantly as $T$ is lowered, $t_0$ increases rapidly as $T$ decreases,  though $\zeta_0$ and $\zeta_R$ only change moderately, from respectively $36.3$ and $107.4$ at $T = 0.8T_c$ to $59.9$ and $80.5$ at $T \approx 0.6T_c$, with reverse temperature dependences as Fig.~\ref{trt0} displays. Therefore, the transition occur at a large field and the hysteresis goes up as $T$ is lowered, as can also be seen in Figs.~\ref{rgt}(b) and (f).

\begin{figure}[b]
\centerline{\includegraphics[width=\linewidth]{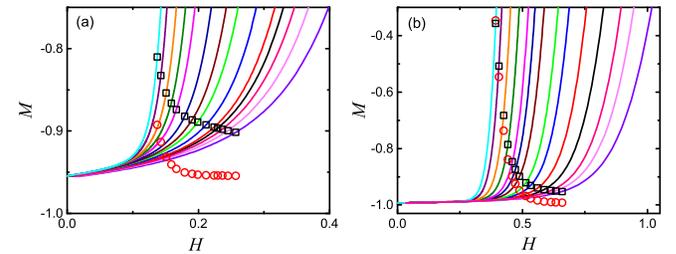}}
\caption{\label{hsms}(Color online) The instability points and the magnetization at Hs for (a) $T = 0.8T_c$ and (b) for $T \approx 0.6T_c$. $M(H_s)$ (squares) lies on the magnetization curves (lines) whereas $M_s$ (circles) does not. In (a) and (b), the field rates $R$ increase from left to right. Three more largest rates are drawn for both temperatures as compared to Figs.~\ref{rgt}(b) and (f). Two more smallest rates are also plotted in (b). The color codes are identical with Fig.~\ref{rgt}. Note that the instability point of the smallest rate in (b) exceeds the magnetization at $H_s$.}
\end{figure}
Figure~\ref{hsms} illustrates the magnetization at $H_s$, $M(H_s)$, and the instability points $(H_s, M_s)$. Their differences are just $f(0)R^{\beta/r\nu}$ from Eq.~(\ref{ftsm}). One sees that scaling and universality persist though the instability points appear somehow far away. It is clear that $H_s$ decreases while $M_s$ increases as $R$ is reduced as expected. A unique feature is that, for the low temperature, $M(H_s)$ and $M_s$ rise sharply for low $R$ and thus low $H_s$ and cross each other. We believe this is the reason why the small rates deviate from scaling at the temperature shown in Fig.~\ref{rgt}(e). There are two possible causes. One is that the nucleation barrier is large and the transition fluctuates a lot. Even though we have $20$ million samples, the average magnetization may still have sufficient fluctuations. Another possible limitation to the accuracy of $M$ may come from finite size effects. Anyway, such a small shift in the $H$ position can thus give rise to a large deviation in $M(H_s)$ in comparison with the theoretical $H_s$. For the large rates, their magnetizations may be too nonequilibrium to show scaling. Indeed, they are found to depend slightly on the initial state.

In addition, owing to the difference between $M(H_s)$ and $M_s$, the time scale coefficient we have found is different from the value $5.59$ obtained from its definition at $T = 0.8T_c$. Conversely, the present value of $36.3$ leads back to $B = 0.000 092 0$, about $270$ times smaller as compared to $0.025 15$. This appears not so absurd as deviations of orders of magnitude are common in the field. For example, to fit the results at the same temperature in the single droplet regime, $B$ must be more than $2 000$ times bigger~\cite{Zhongc}.

We have constructed and verified a theory for the dynamics of first-order phase transitions by integrating the theory of nucleation and growth with the $\phi^3$ RG theory for dynamic scaling and universality in first-order phase transitions and the theory of finite-time scaling.  The theory relies on the time scale of nucleation and growth and the time scale of driving and offers a new physical interpretation of the instability points and different regimes in the dynamics. On the one hand, despite being interwoven with nonuniversal nucleation and growth, scaling and universality have been unambiguously verified in the 2D Ising model below its critical temperature. As a consequence, first-order phase transitions can be studied similar to their continuous counterpart and the theories for both kinds of transitions can be unified. On the other hand, the intimate relationship with scaling and universality provides a new way to accurately determine nucleation and growth.

\begin{acknowledgments}
I thank Xuanmin Cao and Weilun Yuan for their useful discussions. This work was supported by National Natural Science Foundation of China (Grant No. 11575297).
\end{acknowledgments}

\end{document}